\documentclass[aps,prl,twocolumn,superscriptaddress, showpacs]{revtex4}
\usepackage{graphicx}
\usepackage{amsmath,amssymb}
\newcommand{\bm}[1]{\mbox{\boldmath{$#1$}}}
\begin{document}

\title{Kondo Phase Transitions of Magnetic Impurities in Carbon Nanotubes}
\author{Tie-Feng Fang}
\affiliation{Center for Interdisciplinary Studies, Lanzhou University, Lanzhou 730000, China}
\affiliation{Institute of Physics, Chinese Academy of Sciences, Beijing 100080, China}
\author{Qing-feng Sun}
\affiliation{Institute of Physics, Chinese Academy of Sciences, Beijing 100080, China}

\begin{abstract}
We propose carbon nanotubes (CNTs) with magnetic impurities as a versatile platform to achieve unconventional Kondo physics, where the CNT bath is gapped by the spin-orbit interaction and surface curvature. While the strong-coupling phase is inaccessible for the special case of half-filled impurities in neutral armchair CNTs, the system in general can undergo quantum phase transitions to the Kondo ground state. The resultant position-specific phase diagrams are investigated upon variation of the CNT radius, chirality, and carrier doping, revealing several striking features, e.g., the existence of a maximal radius for nonarmchair CNTs to realize phase transitions, and an interference-induced suppression of the Kondo screening. We show that by tuning the Fermi energy via electrostatic gating, the quantum critical region can be experimentally accessed.
\end{abstract}
\maketitle

Carbon nanotubes (CNTs) are formed by wrapping a graphene sheet into a cylinder of nanometer radius \cite{Charlier2007}. Their exceptional electronic structure has allowed the exploration of various fascinating Kondo phenomena, including the singlet-triplet Kondo resonance \cite{Nygard2000}, the enhanced shot noise \cite{Delattre2009} due to the $SU(4)$ Kondo effect \cite{Choi2005}, and the competitions with ferromagnetism \cite{Hauptmann2008} as well as superconductivity \cite{Buitelaar2002}. These studies have utilized short CNTs to construct quantum dots behaving as artificial magnetic impurities. Long CNTs, on the other hand, can play the role of one-dimensional (1D) host for a real magnetic impurity, which may be either a magnetic adatom on the top of a carbon atom ($T$ site) or at the center of a hexagon ($C$ site), or a substitutional dopant in a carbon vacancy ($S$ site). Indeed, the Kondo effect for cobalt clusters adsorbed on metallic CNTs has already been observed \cite{Odom2000}. This has spurred several theoretical works \cite{Clougherty2003,Fiete2002,Baruselli2012} to address related issues. However, a generic Kondo model of a CNT-hosted magnetic impurity, pertaining to arbitrary positions at the atomic scale, has not yet been established. More importantly, while these theories have all considered the metallic-CNT host as 1D normal metal to yield conventional Kondo physics \cite{Hewson1993}, recent experiments \cite{Kuemmeth2008} and calculations \cite{Huertas-Hernando2006,Klinovaja2011,Valle2011} have demonstrated that metallic CNTs can not be normal metal, always having rich low-energy band structures due to the spin-orbit interaction (SOI) and curvature effect. Interesting Kondo physics then arises when these ingredients in the CNT host are included.

In this paper, after establishing a generic Hamiltonian for magnetic impurities in metallic CNTs, we show that depending on explicit impurity positions, the system can maps onto two kinds of hard-gap Kondo models whose host density of states (DOS) are identically gapped by the SOI and curvature effect, but scale distinctly outside the gap region due to the quantum interference between different hybridization paths. We combine renormalization group (RG) arguments and slave boson (SB) techniques to demonstrate that the local-moment (LM) state persists for half-filled impurities in neutral armchair CNTs due to the particle-hole (p-h) symmetry. Away from this special case, quantum phase transitions (QPTs) exist in these gapped systems, separating the Kondo and LM ground states. The resultant phase diagrams are characterized by the CNT radius, chirality, carrier doping, and the impurity position. The interference is found to reduce the Kondo regime, making $S$/$C$ configurations unfavorable for Kondo screening as compared with $T$ sites. For sufficiently deep impurity levels, two quantum critical regions are accessible by scanning tunneling probes and gating the CNT host, with signatures characterizing the nonarmchair from armchair chilarities.

Our starting point is the Anderson Hamiltonian of a magnetic impurity in graphene, $\mathcal{H}\hspace{-0.10cm}=\hspace{-0.10cm}\mathcal{H}_o +\mathcal{H}_g+\mathcal{H}_{og}$, where $\mathcal{H}_o\hspace{-0.10cm}=\hspace{-0.10cm}\sum_\sigma \varepsilon_d d^\dagger_\sigma d_\sigma+\frac{U}{2}\sum_\sigma d^\dagger_\sigma d_\sigma d^\dagger_{\bar{\sigma}} d_{\bar{\sigma}}$ models the impurity as usual \cite{Hewson1993}. $\mathcal{H}_g$ is the graphene tight-binding Hamiltonian reading $\mathcal{H}_g\hspace{-0.10cm}=\hspace{-0.10cm}\sum_{\left\langle i,j\right\rangle,\sigma}t a_{\sigma}^{\dagger
}(\mathbf{R}_{ai})b_{\sigma}(\mathbf{R}_{bj})+\text{H.c.}$, here $a_{\sigma}(\mathbf{R}_{ai})$ [$b_{\sigma}(\mathbf{R}_{bj})$] annihilates an $\pi$-band electron on sublattice A (B) at position $\mathbf{R}_{ai}$ ($\mathbf{R}_{bj}$), and $t$ is the nearest-neighbor hopping energy. The hybridization term $\mathcal{H}_{og}\hspace{-0.10cm}=\hspace{-0.10cm}\sum_\sigma g_\sigma^\dagger d_\sigma + \text{H.c.}$ with $g^\dagger_\sigma\hspace{-0.10cm}=\hspace{-0.10cm}\sum_{\left\langle j \right\rangle}\big[V_{aj}a^\dagger_\sigma(\mathbf{R}_{aj}) + V_{bj}b^\dagger_\sigma(\mathbf{R}_{bj})\big]$, where $\left\langle j \right\rangle$ stands for the A and/or B sublattice sites nearest to the impurity, and $V_{xj}$ ($x\hspace{-0.10cm}=\hspace{-0.10cm}a,b$) represent the corresponding hybridization amplitudes. In particular, $g^\dagger_\sigma\hspace{-0.10cm}=\hspace{-0.10cm}V_{a1} a^\dagger_\sigma(0)$ for a $T$-site adatom. $C$-site impurities can hybridize with six surrounding carbon atoms, yielding $g^\dagger_\sigma\hspace{-0.10cm}=\hspace{-0.10cm}\sum_{j=1}^3 \big[V_{aj}a^\dagger_\sigma(\mathbf{X}_{j}) + V_{bj}b^\dagger_\sigma(-\mathbf{X}_{j})\big]$ with $\mathbf{X}_{j}$ the lattice nearest-neighbor vectors, while $g^\dagger_\sigma\hspace{-0.10cm}=\hspace{-0.10cm}\sum_{j=1}^3 V_{bj}b^\dagger_\sigma(\mathbf{X}_{j})$ for $S$-site impurities on sublattice A. In momentum space, the fermionic basis $c_{\mathbf{k}\sigma\alpha}\hspace{-0.10cm}\equiv\hspace{-0.10cm}\frac{1}{\sqrt{2}}(\alpha a_{\mathbf{k}\sigma}+\frac{\phi_{\mathbf{k}}}{|\phi_\mathbf{k}|}b_{\mathbf{k}\sigma})$ diagonalizes the graphene Hamiltonian as $\mathcal{H}_g\hspace{-0.10cm}=\hspace{-0.10cm}\sum_{\mathbf{k}, \sigma,\alpha}\varepsilon_{\alpha}(\mathbf{k}) c^\dagger_{\mathbf{k}\sigma\alpha}c_{\mathbf{k}\sigma\alpha}$,
where $\phi_\mathbf{k}\hspace{-0.10cm}=\hspace{-0.10cm}\sum_{j=1}^3 e^{i\mathbf{k}\cdot\mathbf{X}_{j}}$,
$\varepsilon_{\alpha}(\mathbf{k})\hspace{-0.10cm}=\hspace{-0.10cm}\alpha t|\phi_\mathbf{k}|$, and $\alpha\hspace{-0.10cm}=\hspace{-0.10cm}\pm1$. Close to the Dirac points $\mathbf{K}$, the dispersion is linear, i.e., $\varepsilon_{\alpha}(\mathbf{K}+{\bm\kappa})\hspace{-0.10cm} \simeq\hspace{-0.10cm}\alpha\hbar v_F|{\bm\kappa}|$ for $|{\bm\kappa}|\hspace{-0.10cm}\ll\hspace{-0.10cm}|\mathbf{K}|$, with $v_F$ the Fermi velocity.
In this basis, the hybridization becomes
$\mathcal{H}_{og}\hspace{-0.10cm}=\hspace{-0.10cm} \sum_{\mathbf{k},\sigma,\alpha} V_{\alpha}(\mathbf{k})c^\dagger_{\mathbf{k}\sigma\alpha}d_\sigma +\text{H.c.}$, where $V_{\alpha}(\mathbf{k})\hspace{-0.10cm}=\hspace{-0.10cm}(\alpha \Phi_{a\mathbf{k}}+\Phi_{b\mathbf{k}}|\phi_\mathbf{k}|/\phi_\mathbf{k}^\ast)/\sqrt{2N}$, with $\Phi_{x\mathbf{k}}\hspace{-0.10cm}=\hspace{-0.10cm}
\sum_{\left\langle j \right\rangle}V_{xj}e^{-i\mathbf{k}\cdot\mathbf{R}_{xj}}$ and $N$ the number of sublattice sites.

We now roll up the graphene sheet along the chiral vector $\mathbf{C}_h\hspace{-0.10cm}=\hspace{-0.10cm}n_1\mathbf{a}_1 +n_2\mathbf{a}_2$ to create a ($n_1$,\,$n_2$) CNT \cite{Charlier2007}, where $n_1,n_2\in\mathbb{Z}$ and $\mathbf{a}_1,\mathbf{a}_2$ are the primitive lattice vectors. While ${\bm\kappa}$'s component parallel to the tube axis, $p\hspace{-0.10cm}\equiv\hspace{-0.10cm}\kappa_\parallel$, remains continuous for CNTs of long length $L$, the periodic boundary condition, $(\mathbf{K}+{\bm\kappa})\cdot\mathbf{C}_h\hspace{-0.10cm} =\hspace{-0.10cm}2\pi m$, $m\in\mathbb{Z}$, quantizes {\bm\kappa}'s perpendicular component, $q_\tau\hspace{-0.10cm}\equiv\hspace{-0.10cm}\kappa_\bot \hspace{-0.10cm}=\hspace{-0.10cm}(m+\tau\nu/3)/R$. Here $R\hspace{-0.10cm}=\hspace{-0.10cm}\frac{a}{2\pi}\sqrt{n^2_1 +n^2_2+n_1n_2}$ is the tube radius with $a\hspace{-0.10cm}=\hspace{-0.10cm}|\mathbf{a}_1|\hspace{-0.10cm} \simeq\hspace{-0.10cm}2.46${\AA} the lattice constant, the valley index $\tau\hspace{-0.10cm}=\hspace{-0.10cm}\pm1$ denotes the two inequivalent $\mathbf{K}^+,\mathbf{K}^-$ Dirac points, and $\nu\hspace{-0.10cm}=\hspace{-0.10cm}\textrm{mod}(n_1-n_2,3)$ characterizes the metallic ($\nu\hspace{-0.10cm}=\hspace{-0.10cm}0$) or semiconducting ($\nu\hspace{-0.10cm}=\hspace{-0.10cm}\pm1$) CNTs. Restricting the graphene quantities, $\varepsilon_\alpha(\mathbf{k}),\,V_{\alpha}(\mathbf{k}),\,c_{\mathbf{k}\sigma\alpha}$, only to these allowed Bloch states near the Dirac points yields corresponding quantities for the CNT: the $\pi$-band spectrum $\varepsilon_{p\tau\alpha}\hspace{-0.10cm}=\hspace{-0.10cm} \alpha\hbar v_F\sqrt{p^2+q_\tau^2}$, the hybridization $V_{p\tau\alpha}\hspace{-0.10cm}=\hspace{-0.10cm} V_\alpha(\mathbf{K}^\tau+{\bm\kappa})\big|_{{\bm\kappa}= (p,\,q_\tau)}$, and the operator $c_{p\sigma\tau\alpha}$.

The surface curvature of CNTs induces the $\pi$ band hybridizing with other high-energy bands (CIH), and enhances the effect of intrinsic SOI, $V_{\text{so}}$, of carbon atoms. At second order in perturbation theory based on a double expansion of $V_{\text{so}}$ and $a/R$ \cite{Klinovaja2011,Valle2011}, the SOI gives a spin-dependent shift $\frac{\sigma\alpha_1V_{\text{so}}a}{\hbar v_FR}$ of $q_\tau$ and directly shifts the energy dispersion by $-\sigma\tau\alpha_2V_{\text{so}}(a/R)\cos3\theta$, while the CIH only causes a valley-dependent $q_\tau$ shift $\frac{\tau\beta a^2\cos3\theta}{\hbar v_FR^2}$. Here, the spin $\sigma\hspace{-0.10cm}=\hspace{-0.10cm}\pm$, the parameters, $\alpha_1$, $\alpha_2$, $\beta$, relate to some unperturbed hopping amplitudes between carbon orbitals \cite{SM}, and $\theta$ is the angle between the chiral vector and the zigzag direction along $\mathbf{a}_1$. Due to the hexagonal symmetry, this chiral angle is constrained to $0\hspace{-0.10cm}\leqslant\hspace{-0.10cm}\theta\hspace{-0.10cm} \leqslant\hspace{-0.10cm}30^\circ$, as calculated by $\theta\hspace{-0.10cm}=\hspace{-0.10cm}\arctan\frac{\sqrt{3}n_2} {2n_1+n_2}$ for $0\hspace{-0.10cm}\leqslant\hspace{-0.10cm}n_2\hspace{-0.10cm} \leqslant\hspace{-0.10cm}n_1$ only. These corrections result in $q_\tau\to q_{\sigma\tau}\hspace{-0.10cm}=\hspace{-0.10cm}\frac{\tau \Delta_{\text{cv}}}{\hbar v_F}+\frac{\sigma \Delta_{
\text{so1}}}{\hbar v_F}$ for the lowest metallic $\pi$ subband,
$\varepsilon_{p\tau\alpha}\to \varepsilon_{p\sigma\tau\alpha}\hspace{-0.10cm}= \hspace{-0.10cm}\alpha\hbar v_F\sqrt{p^2+q_{\sigma\tau}^2}-\sigma\tau \Delta_{\text{so2}}$, and
$V_{p\tau\alpha}\to V_{p\sigma\tau\alpha}\hspace{-0.10cm}=\hspace{-0.10cm} V_\alpha(\mathbf{K}^\tau+{\bm\kappa} )\big|_{\bm\kappa=(p,\,q_{\sigma\tau})}$, by setting $\Delta_{\text{so}1}\hspace{-0.10cm}=\hspace{-0.10cm}\alpha_1 V_{\text{so}}a/R$, $\Delta_{\text{so}2}\hspace{-0.10cm}=\hspace{-0.10cm}\alpha_2 V_{\text{so}}(a/R)\cos3\theta$, and $\Delta_{\text{cv}}\hspace{-0.10cm}=\hspace{-0.10cm}\beta(a/R)^2\cos3\theta$. Our generic Anderson Hamiltonian for a magnetic impurity coupled to the metallic CNT host then reads $\mathcal{H}\hspace{-0.10cm}=\hspace{-0.10cm}\mathcal{H}_o + \mathcal{H}_{c}$,
\begin{eqnarray}
\mathcal{H}_c=\hspace{-0.20cm}\sum_{p,\sigma,\tau,\alpha} \hspace{-0.20cm}\big[\varepsilon_{p\sigma\tau\alpha} c^\dagger_{p\sigma\tau\alpha}c_{p\sigma\tau\alpha}
\hspace{-0.10cm}+\hspace{-0.10cm}\big(V_{p\sigma\tau\alpha} c^\dagger_{p\sigma\tau\alpha}d_\sigma\hspace{-0.10cm}+ \hspace{-0.10cm}\text{H.c.}\big)\big],
\end{eqnarray}
with the host DOS, $\rho_{\sigma\tau}(\varepsilon)\hspace{-0.10cm}\equiv \hspace{-0.10cm}\sum_{p,\alpha}\delta(\varepsilon - \varepsilon_{p\sigma\tau\alpha})$, given by
\begin{equation}
\rho_{\sigma\tau}(\varepsilon)=\rho_{_0}\frac{\big|\varepsilon +\sigma\tau\Delta_{\text{so}2}\big|\Theta\big(\big|\varepsilon +\sigma\tau\Delta_{\text{so}2}\big|-\Delta_{\sigma\tau}\big)} {\sqrt{(\varepsilon +\sigma\tau\Delta_{\text{so}2})^2-\Delta_{\sigma\tau}^2}},
\end{equation}
where $\rho_{_0}\hspace{-0.10cm}=\hspace{-0.10cm}L/(hv_F)$ and $\Delta_{\sigma\tau}\hspace{-0.10cm}=\hspace{-0.10cm}\big|\Delta_{\text{cv}}+\sigma\tau\Delta_{\text{so1}}\big|$. Note that the BCS-like gap opens even in metallic CNTs.

It is allowed to replace $V_{p\sigma\tau\alpha}$ and $\varepsilon_{p\sigma\tau\alpha}$ in the Hamiltonian (1) by a proper constant coupling $V_0$ and an effective spectrum $\widetilde\varepsilon_{p\sigma\tau\alpha}$, respectively, as long as the resultant effective DOS $\widetilde\rho(\varepsilon)\hspace{-0.10cm}\equiv\hspace{-0.10cm} \sum_{p,\tau,\alpha}\delta(\varepsilon -\widetilde\varepsilon_{p\sigma\tau\alpha})$ satisfies $\widetilde\rho(\varepsilon)\hspace{-0.10cm}=\hspace{-0.10cm} \sum_{p,\tau,\alpha}|V_{p\sigma\tau\alpha}/V_0|^2 \delta(\varepsilon-\varepsilon_{p\sigma\tau\alpha})$ \cite{Bulla2008}. Remarkably, $\widetilde \rho(\varepsilon)$ must be spin-independent due to the nonmagnetic nature of the CNT. By performing a Schrieffer-Wolff transformation \cite{Schrieffer1966}, the system can then be readily mapped onto the Kondo model $\mathcal{H}_K\hspace{-0.10cm}=\hspace{-0.10cm}J\mathbf{\hat S}\cdot\mathbf{\hat s}(0)$ which describes the antiferromagnetic exchange interaction $J\hspace{-0.10cm}=\hspace{-0.10cm}-2V_0^2U/[\widetilde \varepsilon_d(\widetilde\varepsilon_d+U)]\hspace{-0.10cm}>\hspace{-0.10cm}0$ of the impurity spin $\mathbf{\hat S}$ with the host spin $\mathbf{\hat s}(0)$ at the impurity site $0$, where $\widetilde\varepsilon_d\hspace{-0.05cm}=\hspace{-0.05cm} \varepsilon_d-E_F\hspace{-0.05cm}<\hspace{-0.05cm}0$ is the impurity level measured from the Fermi energy $E_F$.

The effective DOS $\widetilde\rho(\varepsilon)$ seen by the impurity is essentially a renormalization of the bare CNT DOS, emerging from the quantum interference between different hybridization paths the electrons can take to hop in and out of the impurity. Hereafter, we focus on a particular class of impurity orbitals that hybridizes equally with the nearest carbon atoms on a given sublattice, i.e., $V_{xj}\hspace{-0.05cm}=\hspace{-0.05cm}V_x$. In this case, when the impurity is located on the $S$ or $C$ site, constructive interference renormalizes the CNT DOS as $\widetilde{\rho}_{_\text{SC}}(\varepsilon)\hspace{-0.10cm} =\hspace{-0.10cm}\sum_\tau(\varepsilon +\sigma\tau\Delta_{\text{so}2})^2\rho_{\sigma\tau} (\varepsilon)/(2Nt^2)$ by defining $V_0\hspace{-0.10cm}=\hspace{-0.10cm}V_b$ for $S$ site or $V_0\hspace{-0.10cm}=\hspace{-0.10cm}\sqrt{V_a^2+V_b^2}$ for $C$ site, whereas $\widetilde\rho_{_\text{T}}(\varepsilon) \hspace{-0.05cm}=\hspace{-0.05cm}\sum_\tau\rho_{\sigma\tau} (\varepsilon)/(2N)$ for $T$-sites adatoms where $V_0\hspace{-0.10cm}=\hspace{-0.10cm}V_a$ and the interference is absent. These DOS represent two distinct classes of hard-gap Kondo models promising for unconventional Kondo physics.

(i) A half-filled ($U=-2\widetilde\varepsilon_d$) impurity coupled to the neutral armchair ($\theta=30^\circ$) CNT, where since $\Delta_{\text{so}2}=\Delta_{\text{cv}}=0$, no CIH effect exists and the SOI opens a gap of width, $2\Delta_{\text{so}1}$, at the Fermi level that exactly crosses the Dirac point. The system then exhibits strict p-h symmetry, which prohibits the appearance of even powers of the local level $\widetilde\varepsilon_d$ in its renormalization by successively integrating out high energy states with energy $\pm\Lambda$ in the band edges \cite{Hewson1993,Fritz2004}. The lowest contribution arising from two-loop vertex renormalizations, up to the leading order in a double expansion of $V_0$ and $\widetilde\varepsilon_d$, gives rise to the RG beta function $\beta(\widetilde\varepsilon_d)=4\widetilde\rho(\Lambda) V_0^2\widetilde\varepsilon_d/\Lambda$. Consequently, the corresponding flow of the Kondo coupling $J=-4V_0^2/\widetilde\varepsilon_d$ reads $\beta(J)=-4\widetilde\rho(\Lambda)V_0^2J/\Lambda$. Solving this RG equation yields
$J(\Lambda)=J(\Lambda_0)\text{exp}\big\{-4V_0^2\int_{\Lambda_0}^{\Lambda}
[\widetilde\rho(\Lambda)/\Lambda^2]\,\text{d}\Lambda\big\}$,
where $\Lambda_0$ is the initial band cutoff. It is evident that consecutive RG transformations increase the effective coupling. However, as the scaling of the gapped Kondo model characterized by $\widetilde\rho_{_\text{SC}}(\varepsilon)$ or $\widetilde\rho_{_\text{T}}(\varepsilon)$, terminates at the gap edge $\Lambda=\Delta_{\text{so}1}$, $J(\Lambda)$ flows to a finite value rather than infinity, signaling the absence of the strong coupling (SC) Kondo phase. Therefore, the impurity ground state is always a local moment, consistent with previous numerical RG calculations on the rectangularly gapped band \cite{Chen1998}.

(ii) An infinite-$U$ impurity in the neutral armchair CNT. Here the p-h symmetry of the whole system is violated, while the CNT bath remains p-h symmetric. RG transformations show that already at one-loop order, vertex renormalization of $\widetilde\varepsilon_d$ and $J$ occurs as the band width is reduced \cite{Hewson1993,Fritz2004}. The resultant RG equation for the dimensionless Kondo coupling $\widetilde{J}=-2\widetilde\rho(\Lambda)V_0^2/\widetilde\varepsilon_d$ scales according to $\beta(\widetilde{J})=[\ln\widetilde\rho(\Lambda)]'\Lambda \widetilde{J}-\widetilde{J}^2$. We solve the beta function as
\begin{equation}
\widetilde J(\Lambda)=\frac{\widetilde\rho(\Lambda)\widetilde J(\Lambda_0)}{\widetilde\rho(\Lambda_0)+\widetilde J(\Lambda_0)\int^\Lambda_{\Lambda_0} [\widetilde\rho(\Lambda)/\Lambda]\,\text{d}\Lambda},
\end{equation}
where the denominator being vanishing or nonvanishing during scaling determines the impurity ground state.

For $T$-site adatoms, the DOS $\widetilde\rho_{_\text{T}}(\varepsilon)=
\widetilde\rho_{_0}\Theta(|\varepsilon|-\Delta_{\text{so}1})
|\varepsilon|/\sqrt{\varepsilon^2-\Delta_{\text{so}1}^2}$ is of BCS-type, with $\widetilde\rho_{_0}\equiv\frac{\rho_{_0}}{N}$. As the scaling proceeds, the denominator of Eq.\,(3) vanishes at the critical band width $\Lambda_c=T^0_{\text{K}}+\Delta^2_{\text{so}1}/4T^0_{\text{K}}$ when $2T^0_{\text{K}}>\Delta_{\text{so}1}$, directing the RG flow towards the SC Kondo fixed point. Here the Kondo temperature $T_{\text{K}}^0\equiv\Lambda_0\exp[-1/\widetilde J(\Lambda_0)]$ is defined as a scaling invariant \cite{Hewson1993} of the normal metallic model [realized by setting $\widetilde\rho(\varepsilon)=\widetilde\rho_{_\text{T}} (\Lambda_0)\simeq\widetilde\rho_{_0}$ for $\Lambda_0\gg\Delta_{\text{so}1}$]. By contrast, for $2T^0_{\text{K}}<\Delta_{\text{so}1}$, or equivalently, $\widetilde J(\Lambda_0)<[\ln{(2\Lambda_0/\Delta_{\text{so}1})}]^{-1}$, the adatom flows to the unscreened LM state since the coupling $\widetilde J(\Lambda)$ already vanishes as the scaling enters into the gap region before it reaches the SC limit.
We thus find a quantum-critical point of the impurity level $\widetilde\varepsilon_{dc}=-2V_0^2\widetilde\rho_{_0} \ln\frac{2\Lambda_0}{\Delta_{\text{so}1}}$ across which, upon lowering $\widetilde\varepsilon_d$, the impurity undergoes a QPT from a screened to an unscreened moment. The explicit $R$ dependence of this phase boundary can be written as $\widetilde\varepsilon_{dc}=\varepsilon_{c1} -2V_0^2\widetilde\rho_{_0}\ln\frac{R}{a}$, with $\varepsilon_{c1}=-2V_0^2\widetilde\rho_{_0} \ln\frac{2\Lambda_0}{\alpha_1V_{\text{so}}}$.

The interference-induced additional scaling $(\varepsilon/t)^2$, imposed on the effective host DOS $\widetilde\rho_{_\text{SC}}(\varepsilon)=
(\varepsilon/t)^2\widetilde\rho_{_\text{T}}(\varepsilon)$ for substitutional dopants or $C$-site adatoms, dramatically changes the RG flow of Eq.\,(3). It features a different phase boundary at $\widetilde\varepsilon_{dc}=\varepsilon_{c2}-f(\frac{R}{a})$ separating the Kondo and LM phases, where $\varepsilon_{c2}=-V_0^2\widetilde\rho_{_0}\frac{\Lambda_0^2}{t^2}$ and $f(\frac{R}{a})=
V_0^2\widetilde\rho_{_0}\frac{\alpha_1^2V_{\text{so}}^2 a^2}{t^2R^2}\ln\frac{2\Lambda_0R}{\alpha_1V_{\text{so}}a}$, by taking $\widetilde\rho_{_\text{SC}}(\Lambda_0)\simeq\widetilde \rho_{_0}\frac{\Lambda_0^2}{t^2}$ for $\Lambda_0\gg\Delta_{\text{so}1}$. For realistic parameters, this boundary is always much shallower than in the $T$-site case, reflecting a reduction of the Kondo regime by the interference.

(iii) An infinite-$U$ impurity coupled to the doped nanotube with arbitrary chiralities. In this general case, the gaps in the two valley sectors of the CNT DOS are different, being centered at $\varepsilon=\pm\Delta_{\text{so}2}$ with width $W_1=2|\Delta_{\text{cv}}-\Delta_{\text{so}1}|$ and $W_2=2(\Delta_{\text{cv}}+\Delta_{\text{so}1})$, respectively. After summation over the two valleys, their overlap constitutes a net gap of width $W=\frac{W_1+W_2}{2}-2\Delta_{\text{so}2}$ centered at $\varepsilon=\text{min}(\Delta_{\text{cv}},\Delta_{\text{so}1}) \equiv\varepsilon_0$, in the effective DOS $\widetilde\rho_{_\text{T}}(\varepsilon)$ and $\widetilde\rho_{_\text{SC}}(\varepsilon)$, provided that $W>0$.

The presence of $\Delta_{\text{so}2}$ and $\Delta_{\text{cv}}$ for the nonarmchair chirality, and the deviation of the Fermi level from the Dirac point in the doped nanotube, definitely break the p-h symmetry of the CNT baths. This renders the previous RG arguments insufficient because all vertex functions will develop structures on a scale $E_F$. The SB technique \cite{Coleman1984} accounts for this complication by introducing an auxiliary boson field to describe the empty state, together with a Lagrange multiplier $\lambda$ to exclude double occupancy in the impurity. The boson field is further condensed to its saddle-point value $r$ which minimizes the free energy, obeying $\frac{2}{\pi}\int_{-\Lambda_0}^{E_F}\text{d}\varepsilon\, \text{Im}[\Sigma(\varepsilon)G(\varepsilon)]=\lambda$,
where $\Sigma(\varepsilon)=\frac{1}{\pi}\int_{-\Lambda_0}^{\Lambda_0} \text{d}\varepsilon^\prime\,
\Gamma(\varepsilon^\prime)(\varepsilon- \varepsilon^\prime+i0^+)^{-1}$, $\Gamma(\varepsilon)=\pi V_0^2\widetilde\rho(\varepsilon)$, and the impurity Green's function $G(\varepsilon)=[\varepsilon-\varepsilon_d -\lambda- r^2\Sigma(\varepsilon)]^{-1}$. After $\lambda$ and $r$ are self-consistently determined \cite{Coleman1984}, the localized level $\varepsilon_d$ is renormalized to a Kondo resonance at $\varepsilon_d+\lambda$. This mean-field treatment correctly describes the Kondo fixed point for impurity levels deep below $E_{F}$, where charge fluctuations are frozen out.

At the critical point, the Kondo temperature, defined as $T_\text{K}\hspace{-0.10cm}=\hspace{-0.10cm}\sqrt{ (\widetilde\varepsilon_d+\lambda)^2+r^4 \Gamma^2(\varepsilon_d+\lambda)}$ \cite{Coleman1984}, must vanish. The saddle-point equation then yields the critical value of the impurity level,
\begin{equation}
\widetilde\varepsilon_{dc}=\frac{1}{\pi}\int_{-\Lambda_0}^{E_F} \text{d}\varepsilon\, \frac{\Gamma(\varepsilon)}{\varepsilon-E_F} +\frac{1}{\pi}\int_{E_F}^{\Lambda_0}\text{d}\varepsilon\,\frac {\Gamma(\varepsilon)}{E_F-\varepsilon}.
\end{equation}
This result is applicable to the doped case as well as to the nonarmchair chirality. Below we present numerical results for realistic CNT parameters \cite{Klinovaja2011,SM}: $\alpha_1\hspace{-0.10cm}=\hspace{-0.10cm}0.055$, $\alpha_2\hspace{-0.10cm}=\hspace{-0.10cm}0.217$, $\beta\hspace{-0.10cm}=\hspace{-0.10cm}93.75\text{meV}$, $V_{\text{so}}\hspace{-0.10cm}=\hspace{-0.10cm}6\text{meV}$, and $t\hspace{-0.10cm}=\hspace{-0.10cm}2.5\text{eV}$.

Figure 1 presents impurity phase diagrams in the ($\widetilde\varepsilon_d$, $R$) plane, when the Fermi energy is tuned to the gap center, $E_F=\varepsilon_0$. The QPTs discussed here exist only if the CNT bath is gapped, such that $\widetilde\rho(E_F)$ vanishes exactly. Solving the inequation $W>0$, we find a critical chiral angle $\theta_0=\frac{1}{3}\arccos(\alpha_1/\alpha_2)\approx25.1^\circ$ and an upper limit of CNT radius $R_0=\beta a/(V_{\text{so}}\alpha_2)\approx17.7\text{nm}$. For $\theta_0<\theta\leqslant30^\circ$, CNTs with arbitrary radius are always gapped ($W>0$), resulting in that the impurity exhibits Kondo and LM ground states separated by transitions at $\widetilde\varepsilon_{dc}$ in the whole range of $R$ [see, e.g., Figs.\,1(a) and 1(b) for the armchair case]. On the other hand, when $0\leqslant\theta\leqslant\theta_0$, the Kondo-LM transition can occur only for $R<R_0$. Beyond this upper limit $R\geqslant R_0$, one has $W\leqslant0$, leaving always a screened impurity state [see, e.g., Figs.\,1(c) and 1(d) for the zigzag case]. The specific $R$-dependence of the phase boundary is also very sensitive to explicit impurity positions, despite that $\widetilde\rho_{_\text{T}}(\varepsilon)$ and $\widetilde\rho_{_\text{SC}}(\varepsilon)$ feature the same gap structure. When the CNT radius increases, the Kondo regime of impurities on $T$ ($S$ or $C$) sites gradually widens (narrows), with the boundary eventually decreasing to $-\infty$ (increasing to $\varepsilon_{c2}$) as $R\to\infty$ for $\theta>\theta_0$ chiralities [see, e.g., Figs.\,1(a) and 1(b)] and $R\to R_0$ for $\theta\leqslant \theta_0$ chiralities [see, e.g., Figs.\,1(c) and 1(d)]. This sensitiveness stems from the quantum interference effect, which dramatically changes the scaling behavior of $\widetilde\rho_{_\text{SC}}(\varepsilon)$ outside the gap region, as compared with $\widetilde\rho_{_\text{T}}(\varepsilon)$. By noting $\varepsilon_{c1}\simeq400\varepsilon_{c2}\simeq-10^8f(1)<0$ for the parameters used here, an interference-induced overall shrinking of the Kondo regime in the $S$/$C$ configurations is also evident, as already emphasized before.

\begin{figure}
\centering
\includegraphics[width=0.8\columnwidth]{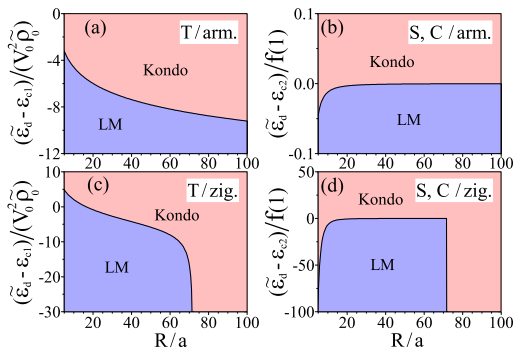}
\caption{(Color online) Phase diagrams of Kondo-LM transitions for impurities sitting on $T$ [(a),(c)], $S$ or $C$ [(b),(d)] sites in the ($n$,$n$) armchair [(a),(b)] and (3n,0) zigzag [(c),(d)] CNTs, as a function of the impurity level $\widetilde\varepsilon_d$ and discrete CNT radius $R/a=\sqrt{3}n/2\pi$ (armchair), $3n/2\pi$ (zigzag), with $n\in\mathbb{Z}$. The high-energy cutoff $\Lambda_0\hspace{-0.10cm}=\hspace{-0.10cm}0.5\text{eV}$.}
\end{figure}

Gating the CNT host to tune its Fermi energy away from the gap center renders $E_F$ closer to electronic states near one of the gap edges, in favor of screening the impurity. Therefore, the SC fixed point can be reached for smaller Kondo couplings (deeper impurity levels), widening the Kondo regime. This is confirmed by our calculations shown in Fig.\,2(a) for $T$-site adatoms \cite{SM}, where the Kondo and LM phases are bounded by an arched borderline peaked at $E_F=\varepsilon_0$. As $E_F$ moves further out of the gap region, arbitrary small $J>0$ can always drive the impurity into the Kondo phase due to $\widetilde\rho(E_F)\neq0$. Interestingly, while the armchair CNT features a p-h symmetric phase diagram when it is tuned from hole doping ($E_F<0$) to electron doping ($E_F>0$), the arched phase boundary of nonarmchair CNTs always deviates from the hole-doped side because of $\varepsilon_0>0$, and can even fully enter into the electron-doped region for large CNT radius. The minimal radius $R_1$ needed for accessing this maximal p-h asymmetry can be determined by solving $\varepsilon_0>\frac{W}{2}$ to obtain $R_{1}=R_0\frac{\alpha_2\cos3\theta} {\alpha_1+\alpha_2\cos3\theta}$ when $\theta\leqslant\theta_0$, thereby $R_1\approx14.1\text{nm}$ for zigzag CNTs.

\begin{figure}
\centering
\includegraphics[width=0.9\columnwidth]{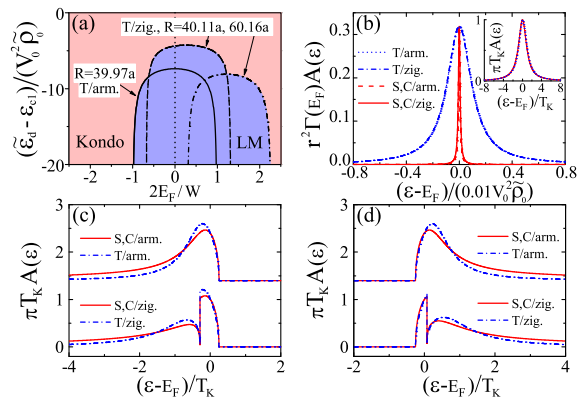}
\caption{(Color online) (a) Phase diagrams for $T$-site adatoms upon variation of the Fermi energy $E_F$, with $\Lambda_0=0.5\text{eV}$. (b)-(d) Spectral densities of the impurity with a deep level $\widetilde\varepsilon_d=-25V^2_0\widetilde\rho_{_0}$ in the ($20$,\,$20$) armchair and ($36$,\,$0$) zigzag CNTs for $E_F=\varepsilon_0-0.88t$ (b), $\varepsilon_0-0.55W$ (c), and $\varepsilon_0+0.55W$ (d), with $\Lambda_0=t$. In (c) and (d), we set $V^2_0=0.1W/[\pi r^2\widetilde\rho(E_F)]$ such that $T_K\approx 0.1W$, and the curves corresponding to the armchair CNT are offset for clarity.}
\end{figure}

Obviously, for deep impurity levels, two consecutive QPTs occur whenever the Fermi energy sweeps over the two gap edges. These can be experimentally observed by scanning tunneling probes \cite{Odom2000} which directly measure the impurity spectral densities, $A(\varepsilon)\hspace{-0.10cm}\equiv\hspace{-0.10cm}-\frac{1}{\pi}\text{Im}G(\varepsilon)$. By placing the Fermi energy far away from the gap region [Fig.\,2(b)], the smooth host DOS around $E_F$ gives rise to conventional Kondo resonances in $A(\varepsilon)$. While the CNT chiralities are indistinguishable in these resonances, the interference inbuilt in the $S$/$C$ configurations greatly narrows the resonances as compared with $T$ sites, signaling a suppression of the Kondo effect which is not favorable for experimental observations, regardless of the existing universal scaling with the Kondo temperature [inset of Fig.\,2(b)]. This scaling is violated when $E_F$ is tuned to access the quantum critical region around the gap edges [Figs.\,2(c) and 2(d)]. Specifically, although the scaling of different impurity positions persists to some extent, the zigzag CNT hosts two-peak Kondo resonances distinct from the armchair one. We attribute the two-peak structure to a distortion by the distinct DOS of nonarmchair CNTs, which possess two singularities around each gap edge, arising from the two valley sectors of the bare DOS of Eq.\,(2). Finally, as $E_F$ shifts into the gap region, the SB equations break down and the Kondo resonances immediately collapse into the featureless LM spectra.

In conclusion, we have addressed the Kondo problem of magnetic impurities in CNTs, demonstrating the existence of distinct QPTs in the impurity's ground state, which crucially depend on the characteristics of CNT and explicit impurity positions.

Support from NBRP of China (2012CB921303 and 2009CB929100) and NSF-China is acknowledged.


\begin{thebibliography}{99}
\bibitem{Charlier2007}
J.-C. Charlier, X. Blase, and S. Roche, Rev. Mod. Phys. \textbf{79}, 677 (2007).
\bibitem{Nygard2000}
J. Nyg{\aa}rd, D.\,H. Cobden, and P.\,E. Lindelof, Nature (London) \textbf{408}, 342 (2000); M. Pustilnik, Y. Avishai, and K. Kikoin, Phys. Rev. Lett. \textbf{84}, 1756 (2000); J. Paaske, A. Rosch, P. W\"{o}lfle. N. Mason, C.\,M. Marcus, and J. Nyg{\aa}rd, Nat. Phys. \textbf{2}, 460 (2006).
\bibitem{Delattre2009}
T. Delattre, C. Feuillet-Palma, L.\,G. Herrmann, P. Morfin, J.-M. Berroir, G. F\`{e}ve, B. Placais, D.\,C. Glattli, M.-S. Choi, C. Mora, and T. Kontos, Nat. Phys. \textbf{5}, 208 (2009); P. Vitushinsky, A.\,A. Clerk, and K.\,Le Hur, Phys. Rev. Lett. \textbf{100}, 036603 (2008); J. Basset, A.\,Yu. Kasumov, C.\,P. Moca, G. Zar\'{a}nd, P. Simon, H. Bouchiat, and R. Deblock, Phys. Rev. Lett. \textbf{108}, 046802 (2012).
\bibitem{Choi2005}
M.-S. Choi, R. L\'{o}pez, and R. Aguado, Phys. Rev. Lett. \textbf{95}, 067204 (2005); P. Jarillo-Herrero, J. Kong, H.\,S.\,J. van der Zant, C. Dekker, L.\,P. Kouwenhoven, and S. De Franceschi, Nature (London) \textbf{434}, 484 (2005); A. Makarovski, J. Liu, and G. Finkelstein, Phys. Rev. Lett. \textbf{99}, 066801 (2007); F.\,B. Anders, D.\,E. Logan, M.\,R. Galpin, and G. Finkelstein, Phys. Rev. Lett. \textbf{100}, 086809 (2008).
\bibitem{Hauptmann2008}
J.\,R. Hauptmann, J. Paaske, and P.\,E. Lindelof, Nat. Phys. \textbf{4}, 373 (2008); M. Gaass, A.\,K. H\"{u}ttel, K. Kang, I. Weymann, J. von Delft, and Ch. Strunk, Phys. Rev. Lett. \textbf{107}, 176808 (2011).
\bibitem{Buitelaar2002}
M.\,R. Buitelaar, T. Nussbaumer, and C. Sch\"{o}nenberger, Phys. Rev. Lett. \textbf{89}, 256801 (2002); A. Eichler, M. Weiss, S. Oberholzer, C. Sch\"{o}nenberger, A.\,L. Yeyati, J.\,C. Cuevas, and A. Mart\'{i}n-Rodero, Phys. Rev. Lett. \textbf{99}, 126602 (2007); J.\,S. Lim, R. L\'{o}pez, and R. Aguado, Phys. Rev. Lett. \textbf{107}, 196801 (2011).
\bibitem{Odom2000}
T.\,W. Odom, J.-L. Huang, C.\,L, Cheung, and C.\,M. Lieber, Science \textbf{290}, 1549 (2000).
\bibitem{Clougherty2003}
D.\,P. Clougherty, Phys. Rev. Lett. \textbf{90}, 035507 (2003).
\bibitem{Fiete2002}
G.\,A. Fiete, G. Zarand, B.\,I. Halperin, and Y. Oreg, Phys. Rev. B \textbf{66}, 024431 (2002).
\bibitem{Baruselli2012}
P.\,P. Baruselli, A. Smogunov, M. Fabrizio, and E. Tosatti, Phys. Rev. Lett. \text{108}, 206807 (2012).
\bibitem{Hewson1993}
A.\,C. Hewson, \textsl{The Kondo Problem to Heavy Fermions} (Cambridge University Press, Cambride, 1993).
\bibitem{Kuemmeth2008}
F. Kuemmeth, S. Ilani, D.\,C. Ralph, and P.\,L. McEuen, Nature (London) \textbf{452}, 448 (2008); H.\,O.\,H. Churchill, F. Kuemmeth, J.\,W. Harlow, A.\,J. Bestwick, E.\,I. Rashba, K. Flensberg, C.\,H. Stwertka, T. Taychatanapat, S.\,K. Watson, and C.\,M. Marcus, Phys. Rev. Lett. \textbf{102}, 166802 (2009); S.\,H. Jhang, M. Marganska, Y. Skourski, D. Preusche, B. Witkamp, M. Grifoni, H. van der Zant, J. Wosnitza, and C. Strunk, Phys. Rev. B \textbf{82}, 041404(R) (2010); T.\,S. Jespersen, K. Grove-Rasmussen, J. Paaske, K. Muraki, T. Fujisawa, J. Nyg{\aa}rd, and K. Flensberg, Nat. Phys. \textbf{7}, 348 (2011).
\bibitem{Huertas-Hernando2006}
D. Huertas-Hernando, F. Guinea, and A. Brataas, Phys. Rev. B \textbf{74}, 155426 (2006); D.\,V. Bulaev, B. Trauzettel, and D. Loss, Phys. Rev. B \textbf{77}, 235301 (2008); L. Chico, M.\,P. L\'{o}pez-Sancho, and M.\,C. Mu\~{n}oz, Phys. Rev. B \textbf{79}, 235423 (2009); J.-S. Jeong and H.-W. Lee, Phys. Rev. B \textbf{80}, 075409 (2009); S. Weiss, E.\,I. Rashba, F. Kuemmeth, H.\,O.\,H. Churchill, and K. Flensberg, Phys. Rev. B \textbf{82}, 165427 (2010).
\bibitem{Klinovaja2011}
J. Klinovaja, M.\,J. Schmidt, B. Braunecker, and D. Loss, Phys. Rev. B \textbf{84}, 085452 (2011).
\bibitem{Valle2011}
M. del Valle, M. Marga\'{n}ska, and M. Grifoni, Phys. Rev. B \textbf{84}, 165427 (2011); W. Izumida, K. Sato, and R. Saito, J. Phys. Soc. Jpn. \textbf{78}, 074707 (2009).
\bibitem{SM}
See Supplemental Material for the explicit forms of the paramters $\alpha_1$, $\alpha_2$, and $\beta$, and for a discussion of the phase diagrams for $S$/$C$ impurities in the ($\widetilde\varepsilon_d$, $E_F$) plane.
\bibitem{Bulla2008}
R. Bulla, T.\,A. Costi, and T. Pruschke, Rev. Mod. Phys. \textbf{80}, 395 (2008).
\bibitem{Schrieffer1966}
J.\,R. Schrieffer and P.\,A. Wolff, Phys. Rev. \textbf{149}, 491 (1966).
\bibitem{Fritz2004}
P.\,W. Anderson, J. Phys. C \textbf{3}, 2436 (1970); F.\,D.\,M. Haldane, Phys. Rev. Lett. \text{40}, 416 (1978); L. Fritz and M. Vojta, Phys. Rev. B \textbf{70}, 214427 (2004).
\bibitem{Chen1998}
K. Chen and C. Jayaprakash, Phys. Rev. B \textbf{57}, 5225 (1998); C.\,P. Moca and A. Roman, Phys. Rev. B \textbf{81}, 235106 (2010).
\bibitem{Coleman1984}
P. Coleman, Phys. Rev. B \textbf{29}, 3035 (1984); \textbf{35}, 5072 (1987); N. Read and D.\,M. Newns, J. Phys. C: Solid State Phys. \textbf{16}, L1055 (1983).


\end{thebibliography}
\end{document}


\renewcommand\theequation{S\arabic{equation}}
\renewcommand\thefigure{S\arabic{figure}}

\title{Supplementary information to
``Kondo phase transitions of magnetic impurities in carbon nanotubes"}
\author{Tie-Feng Fang}
\author{Qing-feng Sun}
\maketitle

\noindent {\bf \large 1) The parameters $\alpha_1$, $\alpha_2$, and $\beta$}

\vspace{3mm}

In the low-energy theory for carbon nanotubes, the effects of spin-orbit interaction and surface curvature on $\pi$ electrons are well described in second-order perturbation theory [14,15]. The effects are equivalent to shift the dispersion relation by $-\sigma\tau\alpha_2 V_{\text{so}}(a/R)\cos3\theta$, to shift the perpendicular wave vector by $\sigma\alpha_1\frac{V_{\text{so}}a}{\hbar v_FR}+\tau\beta\frac{a^2\cos3\theta}{\hbar v_FR^2}$, and to shift the parallel wave vector by $\tau\beta^\prime\frac{a^2\sin3\theta}{\hbar v_FR^2}$. Assuming sufficiently long nanotubes, the last shift is irrelevant, we thus drop it. The explicit forms of the parameters, $\alpha_1$, $\alpha_2$, and $\beta$, appearing in the remaining terms are [14]:
\begin{eqnarray}
\alpha_1&=&-\frac{\sqrt{3}\varepsilon_s(V^\pi_{pp}+V^\sigma_{pp})}{18V^2_{sp}},\\
\alpha_2&=&\frac{\sqrt{3}V^\pi_{pp}}{3(V^\pi_{pp}-V^\sigma_{pp})},\\
\beta&=&\frac{V^\pi_{pp}(V^\pi_{pp}+V^\sigma_{pp})}{8(V^\pi_{pp}-V^\sigma_{pp})}.
\end{eqnarray}
Here $\varepsilon_s$ is the energy of the carbon $s$ orbital relative to the $p$ orbital energy. The latter (i.e., the on-site energy of $\pi$ electrons) is set to zero in our manuscript. $V_{sp}$ represents the unperturbed hopping amplitude between nearest-neighbor $s$ and $p$ orbitals. $V^{\pi(\sigma)}_{pp}$ is the unperturbed hopping amplitude between nearest-neighbor $p$ orbitals, giving rise to the $\pi(\sigma)$ band. In this work, we use the parameter set: $\varepsilon_s=-8.9\text{eV}$, $V_{sp}=5.6\text{eV}$, $V_{pp}^\pi=-3.0\text{eV}$, and $V^\sigma_{pp}=5.0\text{eV}$ [R. Saito, G. Dresselhaus, and M.\,S. Dresselhaus, \textsl{Physical Properties of Carbon Nanotubes} (Imperial College Press, London, 1998)], as it is also used in Ref.\,[14]. This give us, $\alpha_1\simeq0.055$, $\alpha_2\simeq0.217$, and $\beta\simeq93.75\text{meV}$, to carry out numerical calculations. Using other sets of parameters [e.g., D. Tom\'{a}nek and M.\,A. Schluter, Phys. Rev. Lett. \textbf{67}, 2331 (1991); J.\,W. Mintmire and C.\,T. White, Carbon \textbf{33}, 893 (1995)] does not change our numerical results qualitatively.

\vspace{6mm}

\noindent {\bf \large 2) Phase diagrams for $S$- and $C$-site impurities in the ($\widetilde\varepsilon_d$, $E_F$) plane}

\vspace{3mm}

As shown in Fig.\,S1, the Kondo and LM phases of $S$/$C$ impurities are also bounded by an arched borderline, showing features qualitatively same with $T$-site adatoms [see, Fig.\,2(a) in the paper]. For example, the boundary is p-h symmetric for armchair nanotubes, but becomes p-h asymmetric for nonarmchair nanotubes. The minimal radius $R_1$ derived in the paper for accessing the maximal p-h asymmetry also applies to this case. This is because $\widetilde\rho_{\text{sc}}(\varepsilon)$ and $\widetilde\rho_{_\text{T}}(\varepsilon)$ share the same gap structure. They scale differently only outside the gap region due to the quantum interference effect.

The effect of quantum interference is mainly reflected i) in the $R$-dependence of the boundary (see Fig.\,1 in the paper), ii) in the fact that the arched LM region of $S$/$C$ impurities are much sharper than $T$ adatoms [compare Fig.\,2(a) in the paper with Fig.\,S1 here], and iii) in the fact that for realistic nanotube parameters, the Kondo boundary of $S$/$C$ impurities, $\widetilde\varepsilon_{dc}$, is always much shallower than the boundary of $T$ adatoms, signaling the reduction of Kondo regime by interference.
\\
\begin{figure}[H]
\centering
\includegraphics[width=1.0\columnwidth]{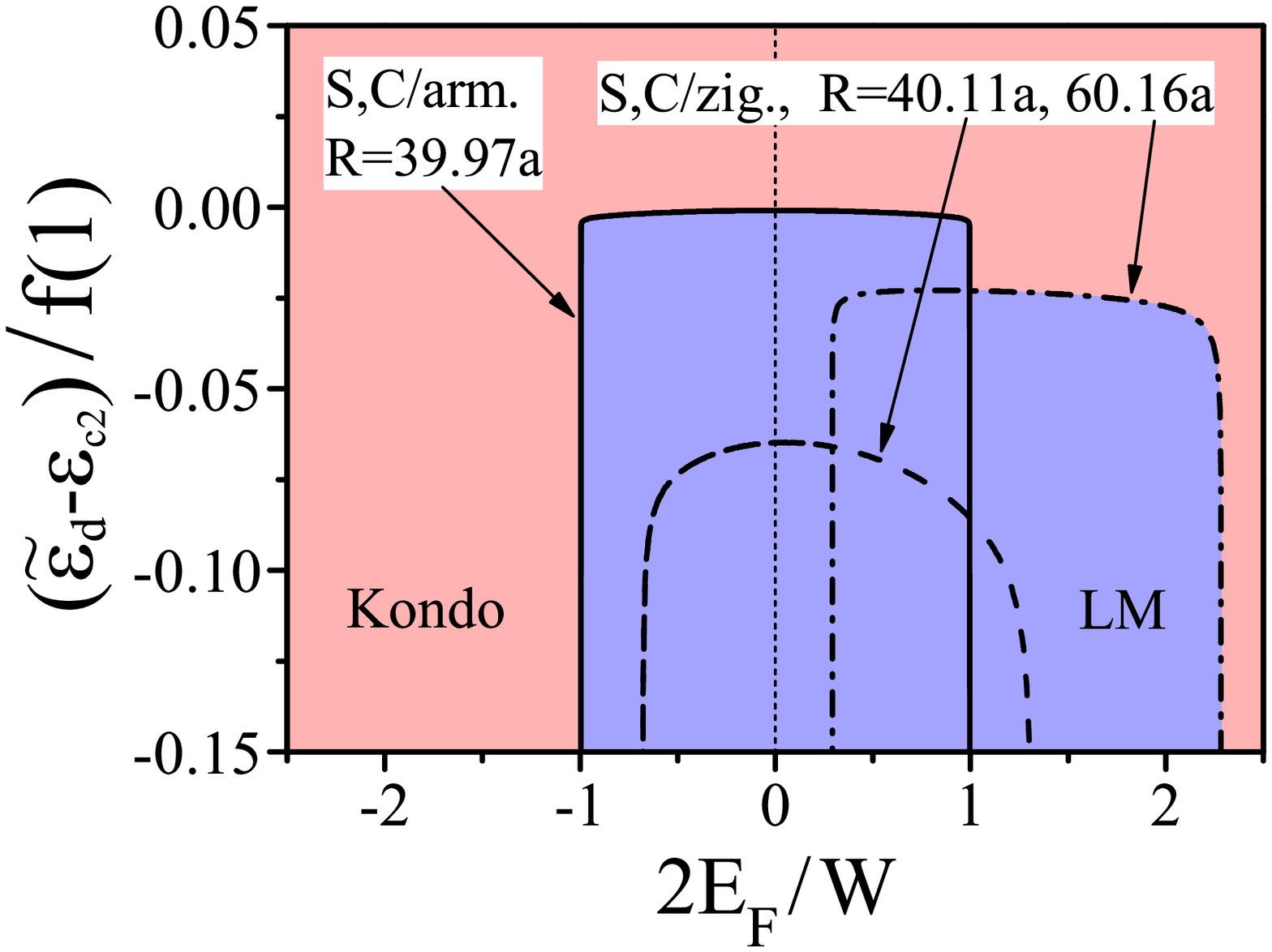}
\caption{Phase diagrams for substitutional dopants or $C$-site adatoms in armchair and zigzag nanotubes in the ($\widetilde\varepsilon_d$, $E_F$) plane. The parameters used are the same as in Fig.\,2(a).}
\end{figure}